\documentclass[%
 reprint,
 amsmath,amssymb,groupedaddress,superscriptaddress,
 prl,longbibliography,floatfix,%preprint,%onecolumn
]{revtex4-2}
\usepackage[T1]{fontenc}
\usepackage{graphicx}
\usepackage{dcolumn}
\usepackage{bm}
\usepackage{amsmath}
\usepackage{lmodern}
\usepackage{mathtools}
\usepackage[colorlinks,citecolor=black,linkcolor=black,urlcolor=black]{hyperref}
\usepackage[normalem]{ulem}
\usepackage{upgreek}

\hypersetup{
    colorlinks,
    linkcolor= [rgb]{0.1804,0.1882,0.5725},
    citecolor=[rgb]{0.1804,0.1882,0.5725},
    urlcolor=[rgb]{0.1804,0.1882,0.5725}
}
\usepackage{chemformula}

\begin{document}
\title{Lasing in the Rashba-Dresselhaus spin-orbit coupling regime in a dye-filled liquid crystal optical microcavity}
\author{Marcin\,Muszy\'nski}
\author{Mateusz\,Kr\'ol}
\author{Katarzyna\,Rechci\'nska}
\author{Przemys\l{}aw\,Oliwa}
\author{Mateusz K\k{e}dziora}
\affiliation{Institute of Experimental Physics, Faculty of Physics, University of Warsaw, Poland}
\author{Rafa\l{}\,Mazur}
\author{Przemys\l{}aw\,Morawiak}
\author{Wiktor\,Piecek}
\affiliation{Institute of Applied Physics, Military University of Technology, Warsaw, Poland}
\author{Przemys\l{}aw\,Kula}
\affiliation{Institute of Chemistry, Military University of Technology, Warsaw, Poland}
\author{Pavlos\,G.\,Lagoudakis}
\affiliation{Skolkovo Institute of Science and Technology, Bolshoy Boulevard 30, bld. 1, Moscow, 121205, Russia}
\affiliation{Department of Physics and Astronomy, University of Southampton, Southampton SO17 1BJ, UK}
\author{Barbara\,Pi\k{e}tka}
\author{Jacek\,Szczytko}
\email{Jacek.Szczytko@fuw.edu.pl}
\affiliation{Institute of Experimental Physics, Faculty of Physics, University of Warsaw, Poland}

\begin{abstract}
In the presence of Rashba-Dresselhaus coupling, strong spin-orbit interactions in liquid crystal optical cavities result in a distinctive spin-split entangled dispersion. Spin coherence between such modes give rise to an optical persistent-spin-helix. In this letter, we introduce optical gain in such a system, by dispersing a molecular dye in a liquid-crystal microcavity. We demonstrate both lasing in the Rashba-Dresselhaus regime and the emergence of an optical persistent spin helix.
\end{abstract}
\maketitle

Engineering the dispersion of photons in microstructured optical systems brings about a plethora of new applications in optoelectronics ranging from information transfer and processing, to quantum optics and beyond \cite{Vahala2003}. Furthermore, it enables the emulation of complex solid-state systems through the realization of synthetic Hamiltonians that can be derived merely from the propagation of electromagnetic waves in appropriately designed photonic microstructures. A compelling example is the photonic analogue of spin-orbit interactions in electronic systems, where the pseudospin of a photon mimics an electron's spin \cite{Rechcinska_Science2019}. The ability to engineer artificial gauge fields for photons \cite{Hey2018} has led to rapid advances in topological photonics \cite{Lu2014,Ozawa_RevModPhys2019} and the realisation of photonic spin-orbit coupling (SOC) Hamiltonians \cite{Sala_PRX2015,Whittaker_NatPhot2020,Bliokh2015} following advancements in cold atoms \cite{Dalibard_RevModPhys2011,Galitski2013,Weitenberg_NatPhys2021} and solids \cite{Basov_NatMater2017,Aidelsburger2018,Rudner_NatRevPhys2020}.

Liquid crystal (LC) birefringent cavities allow for a wide range of tunability of energy and polarization of confined photonic modes \cite{Lekenta_LSA2018,Rechcinska_Science2019}. In particular, if two photonic modes of orthogonal polarization and different parity are tuned to the resonance, a characteristic scheme of spin-polarized dispersion is formed \cite{Rechcinska_Science2019}, which in the solid state physics is called the Rashba-Dresselhaus (RD) spin-orbit coupling \cite{Manchon_NatMater2015}. The characteristic feature of equal Rashba and Dresselhaus terms is the appearance of long-range spin texture, so-called persistent spin helix \cite{Bernevig_PRL2006,Koralek_Nature2009}, recently observed also in photonic system \cite{PSH_arxiv2021}.
Such tunable cavities, which can be easily integrated with various organic \cite{Kuehne_ChemRev2016} and inorganic emitters \cite{Erdem2016,Park_NatRevMater2021}, can provide an excellent platform for photonic and optoelectronic devices. 
There has been tremendous progress in the development of high-efficiency organic semiconductor dyes \cite{Jiang_ChemSocRev2020} which can be also embedded in liquid crystal lasers \cite{Mysliwiec_Nanophot2021}, where dye molecules are simply dissolved in a liquid crystal matrix. 
    In our work we took advantage of the integrated photonic microcavity filled with a liquid crystal of high birefringence \cite{Dabrowski2013} and laser dye pyrromethene 580 (P580) \cite{LpezArbeloa2005,Mowatt2010}.

\begin{figure}%[hbt]
    \centering 
    \includegraphics{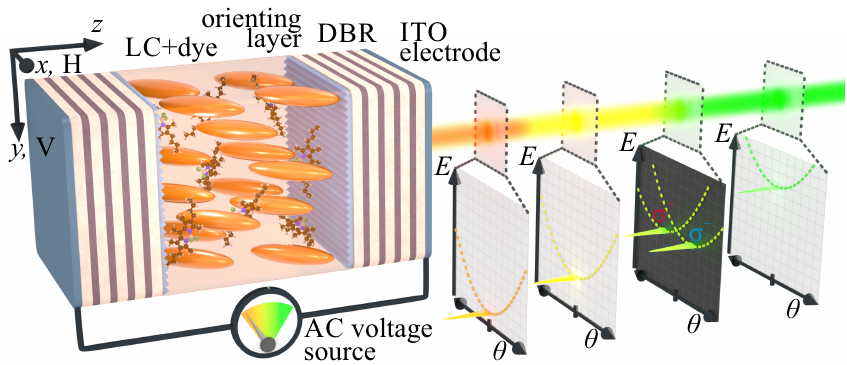} 
    \caption{Scheme of the optical microcavity filled with liquid crystal and P580 dye (dye-doped LCMC). Upon pulsed excitation lasing action occurs at the lowest energy $E$ of the cavity modes dispersion relation, which depends on the amplitude of AC voltage applied to the structure.  Especially at Rashba-Dresselhaus resonance emitted light is split between to two circularly polarized valleys and leaves the cavity at nonzero emission angle $\theta$.}
    \label{im:Fig1}
\end{figure}

A schematic of our device is shown in Fig.~\ref{im:Fig1}. 
The cavity is composed of two $\text{SiO}_2$/$\text{TiO}_2$ distributed Bragg reflectors (DBRs) with maximum reflectance at $\lambda=550$~nm (2.25~eV). Two spacers were placed at the edges of the cavity so that, the space between the DBRs varied from about 2~$\upmu$m to 3~$\upmu$m, depending on the position on the sample. This space was filled with a highly birefringent nematic liquid crystal ($\Delta n= 0.41$). Initial orientation of LC molecules is determined by the polymer orienting layers on top of the DBRs. The use of transparent electrodes made of indium tin oxide (ITO) allows to control the optical anisotropy in the $x$--$z$ plane. The application of an external electric field changes the effective refractive index and thus the energy of the cavity mode for $x$-polarized light (horizontal or H mode), while the energy of $y$-polarized mode (vertical or V mode) remains unchanged.

In the presence of degeneracy between the horizontal and vertical modes with different parity occurs, a Rasha-Dresselhaus spin-orbit coupling interaction appears \cite{Rechcinska_Science2019}. The dispersion relation can then be described by the effective Hamiltonian written in the basis of photon circular polarizations:
\begin{equation} \label{eq:Hexp}
\hat{H}=\frac{\hbar^2{\vec k}^2}{2m}-2\alpha k_y\hat{\sigma}_z + \frac{1}{2}(\Delta E_{\text{HV}}) \hat{\sigma}_x,
\end{equation}
where $\vec{k}=(k_x,k_y)$ is the cavity in-plane momentum, $m$ is the effective mass of
the cavity photon, $\alpha$ is the
Rashba-Dresselhaus coupling coefficient, $\hat{\sigma}_x$ ($\hat{\sigma}_z$) is the first (third) Pauli matrix and $\Delta E_{\text{HV}}$ corresponds to the energy difference between H and V polarized modes.

By doping the cavity with a dye, the spontaneous emission becomes filtered and transformed by cavity modes. Additionally, the dye acts as a gain medium inside the cavity, making it possible to obtain a laser action. The effect of tuning the emission energy with the applied voltage is conceptually presented in Fig.~\ref{im:Fig1}.

Optical measurements were performed at room temperature. All data were acquired in a single shot on demand experiment. The microcavity was excited by a Q-switched diode pumped laser (532~nm center wavelength, 2~ns pulse duration). The microscope objective with numerical aperture $\text{NA} = 0.75$ was used for both excitation and collection of emitted light. The size of the pump spot had a diameter of around 5~$\upmu$m. A quarter-wave plate, half-wave plate and linear polarizer setup was used to detect light in selected polarization. To tune the cavity modes, we used an external voltage of square waveform of frequency 1~kHz.

\begin{figure}%[hbt]
    \centering
    \includegraphics[width=8.6cm]{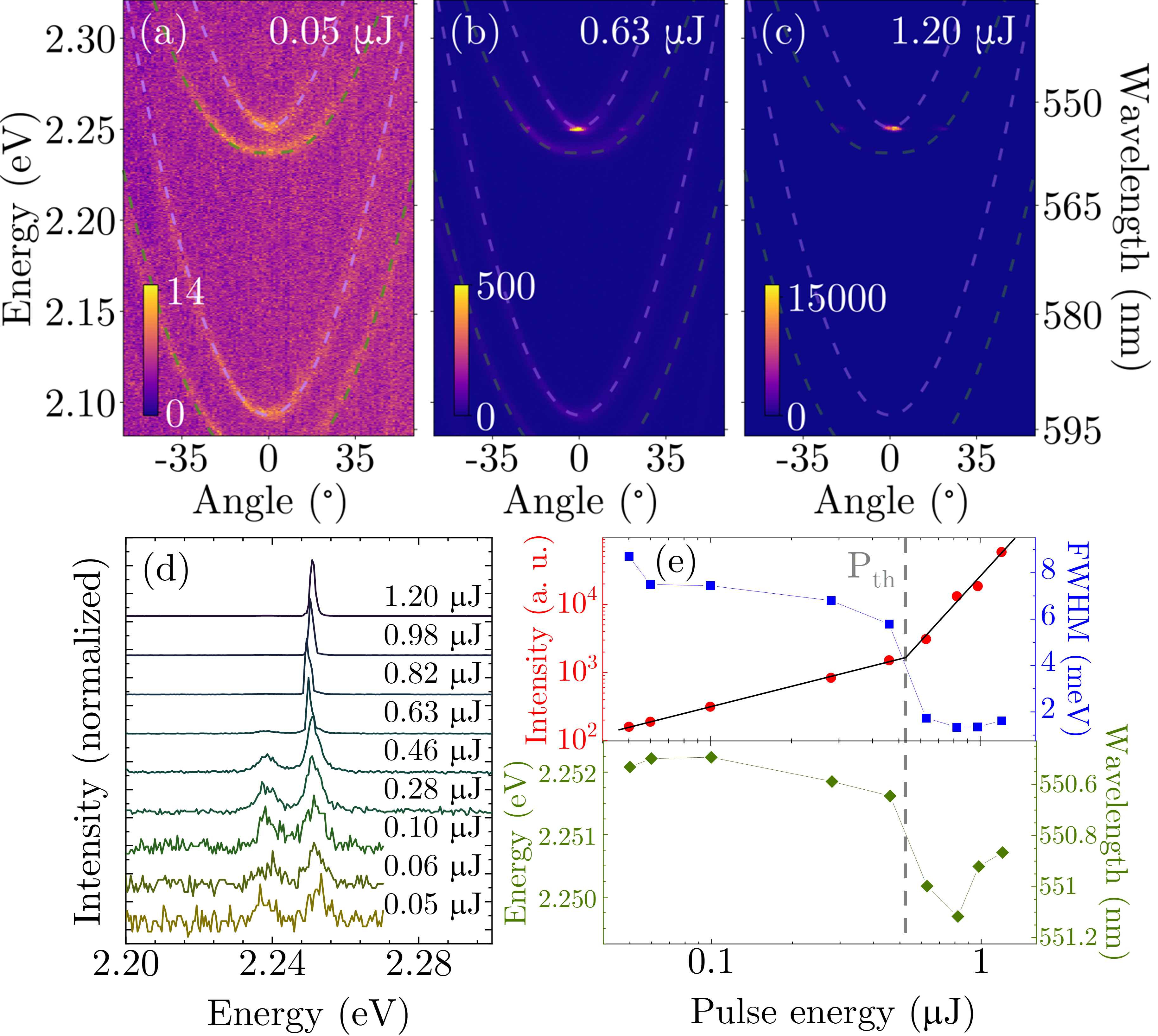} 
    \caption{Lasing in dye-doped LCMC. Experimental emission dispersion (a)~below; (b),(c)~above the lasing threshold. The purple and green dashed lines correspond to the Rashba-Dresselhaus Hamiltonian~\eqref{eq:Hexp}. (d)~Normalized emission spectra obtained for an angle close to 0 degrees for different pulse energies. (e)~Emission intensity (red dots), linewidth (blue squares) and peak position (green diamonds) as a function of the energy of the excitation pulse. The black lines are the linear fits to the intensity parameters below (first 5 points) and above (last 4 points) the lasing threshold. }
    \label{im:Fig2}
\end{figure}

Figure~\ref{im:Fig2} shows the lasing effect in dye-doped liquid crystal microcavity (LCMC). Fig.~\ref{im:Fig2}(a)--\ref{im:Fig2}(c) shows angle-resolved photoluminescence spectrum close to the Rashba-Dresslehaus regime for increasing energies of the excitation pulse. 
In Fig.~\ref{im:Fig2}(a) the emission spectrum obtained below the lasing threshold reveals dispersion relation of the cavity modes. The solutions of Rashba-Dresselhaus Hamiltonian given by Eq.~\eqref{eq:Hexp} were fitted to the experimental results to the two upper  and two lower modes independently, and were plotted on maps with purple and green dashed lines for horizontal and vertical modes, respectively.  
Fig.~\ref{im:Fig2}(b) presents the measurement obtained just above the lasing threshold. At the bottom of the horizontal mode parabola, the enhancement and both spectral and angular narrowing of the light emission takes place. Fig.~\ref{im:Fig2}(c) illustrates the situation well above the threshold, when laser light dominates the entire map. The emission spectra at 0 degrees angle for different pulse energies are plotted in Fig.~\ref{im:Fig2}(d). Measured spectra were fitted with Lorentzian line shape and extracted fitting parameters were shown in Fig.~\ref{im:Fig2}(e). The emission intensity (red dots) initially increases linearly with the excitation power, following the black line fitted to the data points below the lasing threshold. For pulse energy of 0.63~$\upmu$J and above, a rapid, nonlinear increase in the emission intensity is observed and a concurrent linewidth narrowing (blue squares), indicative of lasing threshold. At lasing threshold, the emission wavelength exhibits a redshift followed by a blueshift at higher pumping pulse energies. We note here, that in non-crystalline organic microcavities, wherein excitons are localised on individual chromophores or molecules, wavelength blueshifts of the nonlinear emission with increasing excitation density do not indicate the presence of strong coupling \cite{Yagafarov_CommPhys2020}. The localised nature of excitons precludes repulsive exchange interactions, and the associated blueshifts that are observed with increasing polariton density in crystalline semiconductors. It was recently shown that even in strongly coupled dye-filled microcavities, the observed blueshift at condensation threshold is predominantly driven by a renormalisation of the intracavity refractive index and to a lesser extend by a quenching of the Rabi-splitting \cite{Yagafarov_CommPhys2020}.  It is therefore essential to dissociate claims of strong-coupling with non-linear emission blueshifts at lasing threshold, unless a through characterisation is performed that can preclude mechanisms that are not associated with strong coupling \cite{Yagafarov_CommPhys2020}. Claims of strong coupling should include spectroscopic evidence of anti-crossing, or adequate differentiation with non-strongly coupled reference samples \cite{Li_arxiv2021,Lagoudakis2004}. In this study, we have not identified any evidence of strong coupling and the observed emission wavelength shifts are attributed to excitation density induced changes of the intracavity refractive index including heating.

\begin{figure}%[hbt]
    \centering
    \includegraphics[width=8.6cm]{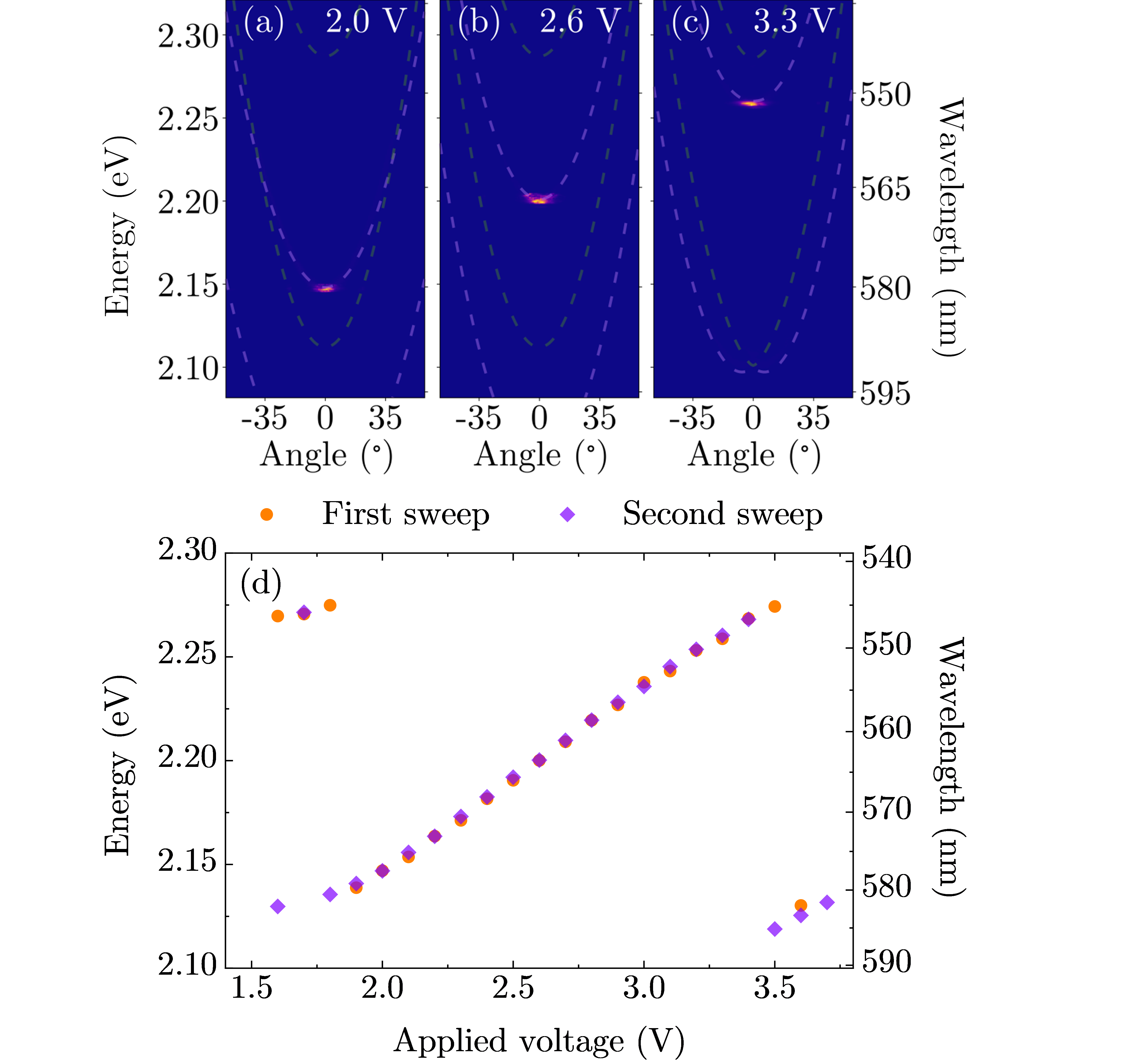} 
    \caption{Tunable laser wavelength. (a)--(c)~Normalized experimental emission dispersion measured for three different applied voltages above the lasing threshold. (d)~Position of the main emission peak versus the applied voltage. The first swipe (orange dots) and the second swipe (purple diamonds) were measured for pulse energies of 0.46~$\upmu$J and 0.63~$\upmu$J, respectively.}
    \label{im:Fig3}
\end{figure}

Since lasing takes place in the horizontal mode -- polarized in the plane of the rotation of LC molecules -- it provides the opportunity to tune the spectral position of the laser line with the applied external voltage. Fig.~\ref{im:Fig3}(a)--\ref{im:Fig3}(c) shows three normalized angle-resolved emission spectra above the lasing threshold for different applied voltages. The lasing line follows the energy of the bottom of the parabola of the horizontal mode. Peak position is shown as a function of the external voltage in Fig.~\ref{im:Fig3}(d). Two voltage sweeps were performed for a pulse energy of 0.46~$\upmu$J (orange dots) and 0.63~$\upmu$J (purple diamonds). The measurements were taken at a different spot on the sample than in Fig.~\ref{im:Fig2}, hence the lasing threshold is lower. One can see that the tuning effect is repeatable and for this position on the sample the available range of peak positions spans approx. 40~nm. This range is limited by dye gain bandwidth and the sample thickness. The second effect is evident for the voltages where the peak position wavelength is above 580~nm. In this case two subsequent horizontal modes begin competing for the population inversion.

\begin{figure}%[hbt]
    \centering
    \includegraphics[width=8.6cm]{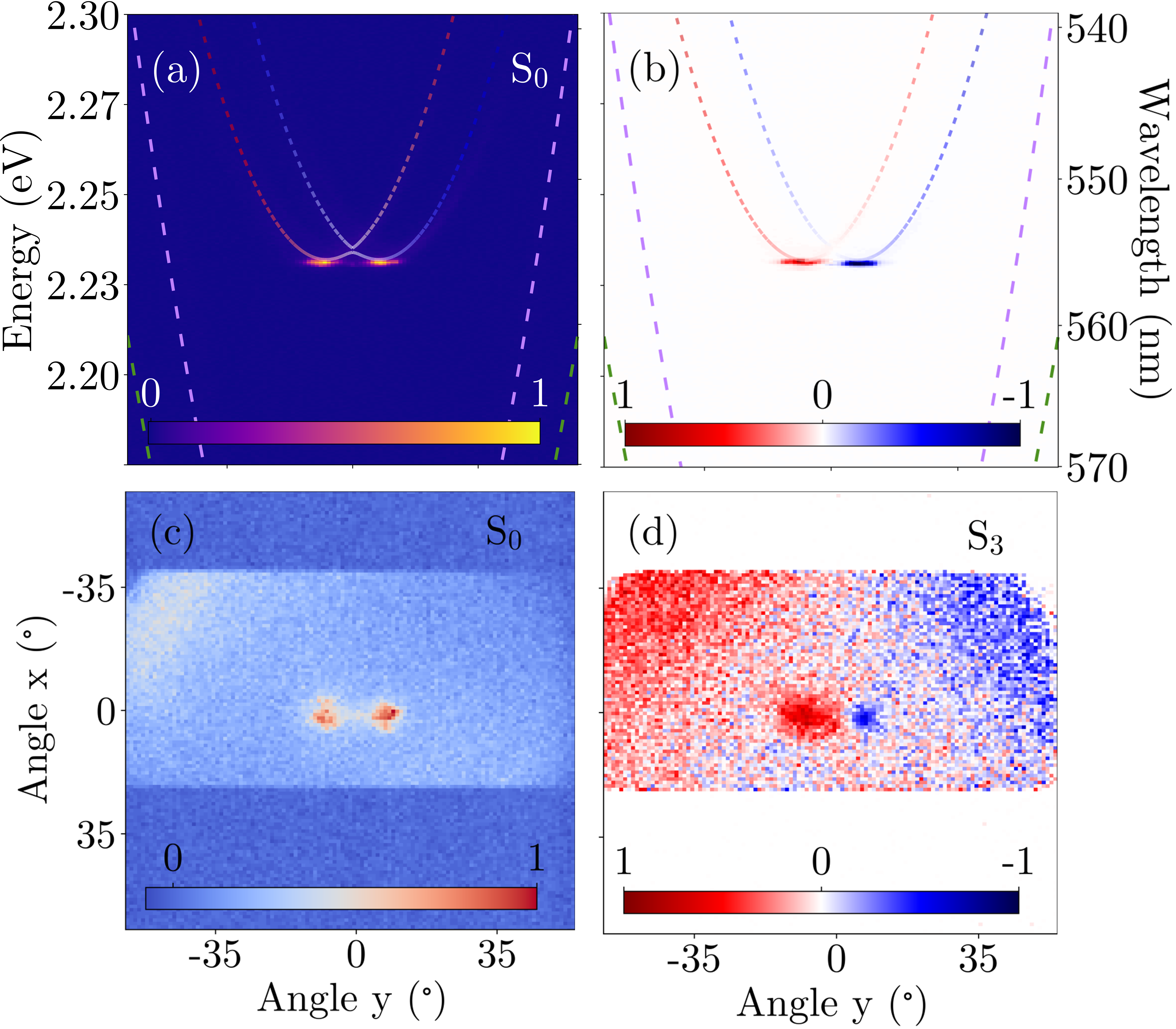} 
    \caption{Lasing in Rashba-Dresselhaus resonance. (a) Angle-resolved emission spectra, (b)~the difference between emission intensities in $\sigma^+$ (red) and $\sigma^-$ (blue) polarization. Momentum space imaging for (c) total emission intensity and (d)~degree of circular polarization ($S_3$ parameter).}
    \label{im:Fig4}
\end{figure}

The interesting effect occurs at the modes degeneracy in the regime of optical spin-orbit Rashba-Dresselhaus (RD) coupling. In this case the lasing takes place simultaneously from the bottoms of two off-centered spin polarized coupled valleys, as shown at angle-resolved emission spectrum in Fig.~\ref{im:Fig4}(a). Since both valleys are circularly polarized with opposite helicities to each other, the laser light also inherits the state of their circular polarization. This effect is evidenced in Fig.~\ref{im:Fig4}(b), presenting the difference of emission intensities in $\sigma^+$ (red) and $\sigma^-$ (blue) polarizations (unnormalized $S_3$ parameter). Fig.~\ref{im:Fig4}(c) and \ref{im:Fig4}(d) presents corresponding results resolved in both axes the of emission angles, but without spectral resolution. In Fig.~\ref{im:Fig4}(c), presenting total emission intensity, one can see two spots spaced apart along the $y$ axis. As expected, both spots are strongly circularly polarized with opposite signs as is shown in Fig.~\ref{im:Fig4}(d) presenting $S_3$ parameter of emitted light. Separation between both spots of $\pm$8$^\circ$ correspond to $Q = 3.1$\,$\upmu$m$^{-1}$ in momentum space as in-plane wave vector is dependent on emission angle $\theta$ by the formula $k_\| = \frac{E}{\hbar c} \sin \theta$. The rectangular shape of the maps in Fig.~\ref{im:Fig4}(c) and \ref{im:Fig4}(d) is related to the spectrometer slit aperture.

The light emitted in RD conditions can be considered as a pair of beams of circular polarizations with equal intensities  given by basis vectors of polarization subspace $\{\vec\sigma_-,\vec\sigma_+\}$ and momentum subspace $\{-\vec{Q} , +\vec{Q} \}$. In the far field it means that determination of direction of light propagation unambiguously determines also its polarization and vice versa. In other words the emission in Fig.~\ref{im:Fig4} is separable into pairs $\{-\vec Q,\vec\sigma_+\}$ and $\{+\vec Q,\vec\sigma_-\}$.  In the near field however the coherent emission from both dispersion spin valleys leads to the interference pattern of linearly polarized persistent spin helix (PSH)  \cite{PSH_arxiv2021}. Such interference requires emission in both circular polarizations so it is given by an inseparable state of light mixing all  basis vectors in momentum and polarization degrees of freedom.
The electric field at the surface of the cavity exactly at the RD resonance conditions is represented by the components of the normalized Stokes vectors \cite{PSH_arxiv2021}: 
\begin{equation} \label{eq.Stokes11}
   \begin{aligned}
     S_1 (\vec r)&= \cos\left(\vec{Q}\cdot\vec{r}+2\Theta\right), \\
     S_2 (\vec r)&= -\sin\left(\vec{Q}\cdot\vec{r}+2\Theta\right), \\
     S_3 (\vec r)&= 0,
   \end{aligned}
   \end{equation}
where $\Theta$ is a variable which uniquely determines the location of the linear polarization interference minima and maxima in the PSH.

\begin{figure}%[hbt]
    \centering
    \includegraphics[width=8.6cm]{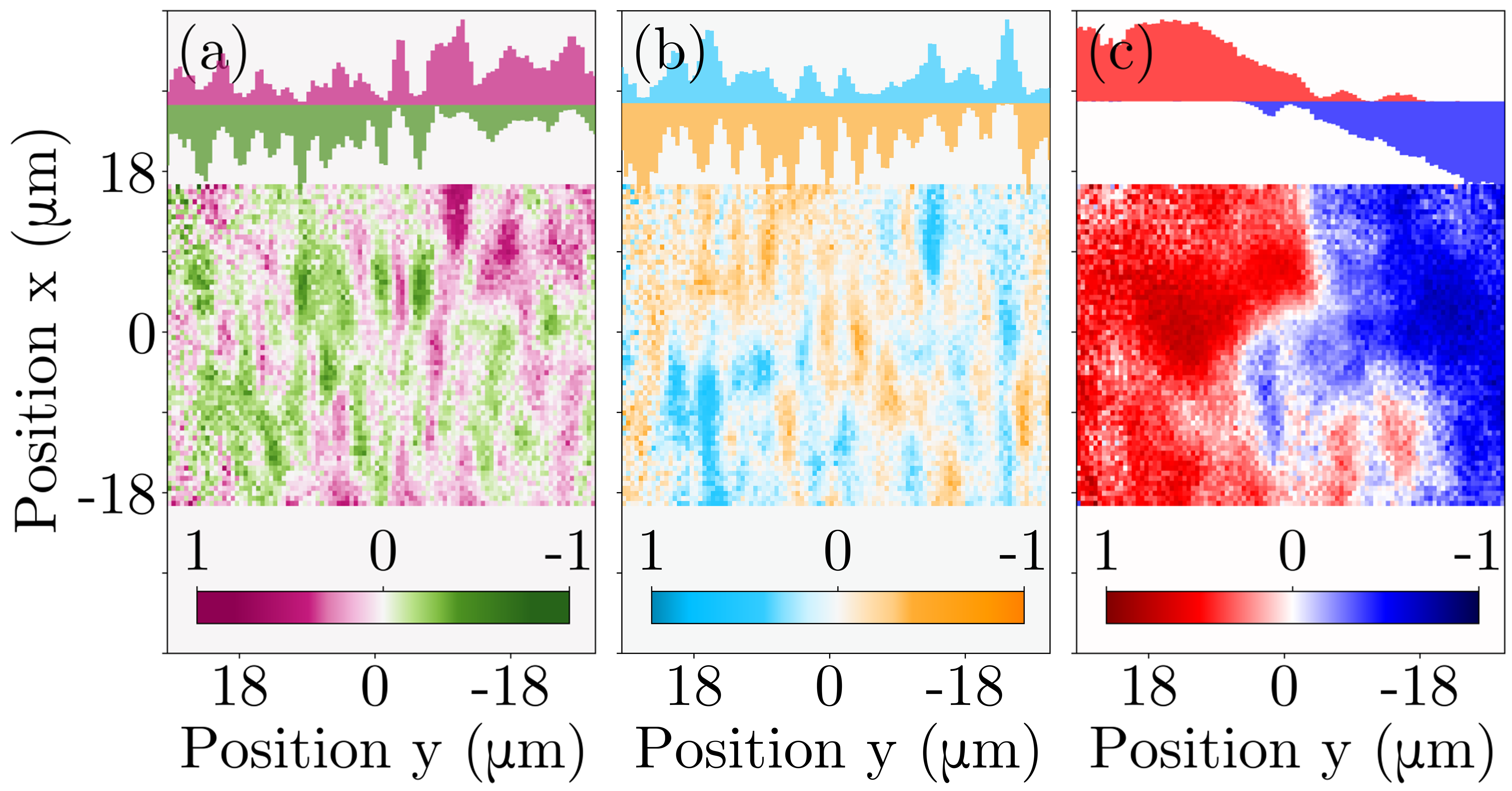} 
    \caption{Persistent spin helix laser. Stokes parameters: (a) $S_1$, (b) $S_2$ and (c) $S_3$ for real-space imaging. The histograms in the insets at the top correspond to the positive and negative values along the $y$-direction.}
    \label{im:Fig5}
\end{figure}

Figure~\ref{im:Fig5} shows a real-space image of Stokes parameters $S_1$, $S_2$ and $S_3$ [Fig.~\ref{im:Fig5}(a), (b), (c), respectively]. The maps of the $S_1(\vec r)$ and $S_2(\vec r)$ parameters show the pattern of longitudinal stripes along the $x$-axis. This proves the spin coherence occurring in the system. To better visualize the PSH pattern, the insets at the top of Fig.~\ref{im:Fig5} shows a histogram made for the positive and negative Stokes parameters values along the $y$-axis. We note that the maxima of the histogram for the horizontal polarization (H) coincide with the minima of the histogram corresponding to the vertical polarization (V) Fig.~\ref{im:Fig5}(a). The helical polarization pattern is further confirmed by comparison of diagonal and antidiagonal polarizations Fig.~\ref{im:Fig5}(b) -- actually according to Eq.~\eqref{eq.Stokes11}  every measurement of two orthogonal linear polarizations projects the order of the polarization (spin) helix and would reveal similar stripes.

Contrary to Eq.~\eqref{eq.Stokes11} the map of  $S_3(\vec r)$ in Fig.~\ref{im:Fig5}(c)  demonstrates a distinct spatial separation of both circular polarizations. The corresponding histogram along $y$-axis reveals increasing absolute value of $S_3(\vec r)$ components with the distance from the excitation spot, where indeed $S_3 (0) \approx 0$. As it has been shown in \cite{PSH_arxiv2021} the spatial distribution of $S_3(\vec r)$, as the consequence of the optical Stern-Gerlach effect, is an useful measure on the amount of inseparability of orbital and polarization degrees of freedom in space. 
For $S_3^2=0$ the system is maximally inseparable whereas when $S_3^2=1$ the spin and valley degrees of freedom are completely separable \cite{PSH_arxiv2021,Korolkova_RPP2019}.
In the center of excitation spot, $S_{3}(0)$ component is close to zero and simultaneously $S_{1}(\vec r)$ and $S_{2}(\vec r)$ both have periodic dependence given by Eq.~\eqref{eq.Stokes11} in Fig.~\ref{im:Fig5}(a) and \ref{im:Fig5}(b). Photons detected far from the excitation spot are mostly circularly polarized -- as expected from angular emission $\pm$8$^\circ$ (or $\pm 0.25$\,$\upmu$m$^{-1}$) shown in Fig.~\ref{im:Fig4}. Thus polarization and direction degrees of freedom are fully separable and $S_3^2 \approx 1$ [Fig.~\ref{im:Fig5}(c)].

In summary, we realized a tunable laser based on a liquid crystal optical microcavity doped with the pyrromethene~580 organic dye. The tunable range reaches 40~nm. Horizontally polarized laser emission is repeatable and can be triggered by single pulses of a pump laser. By transforming the system into the Rashba-Dresselhaus coupling regime, the laser action takes place from the bottoms of two oppositely polarized valleys shifted apart in reciprocal space. Measurements of emission in real space showed the presence of persistent spin helix polarization texture, which is evidence of the spin coherence of the system. The platform we proposed can be used in quantum communication, in which information is encoded through light polarization.

This work was supported by the National Science Centre grants 2019/35/B/ST3/04147, 2019/33/B/ST5/02658 and 2017/27/B/ST3/00271, and the Ministry of National Defense Republic of Poland Program -- Research Grant MUT Project 13-995 and MUT University grant (UGB) for the Laboratory of Crystals Physics and Technology for year 2021. P.G.L. acknowledge the support of the UK's Engineering and Physical Sciences Research Council (grant EP/M025330/1 on Hybrid Polaritonics), the support of the RFBR project No. 20-52-12026 (jointly with DFG) and No. 20-02-00919, and the European Union's Horizon 2020 program, through a FET Open research and innovation action under the grant agreements No. 899141 (PoLLoC) and No. 964770 (TopoLight).

%\bibliography{bib}

%

\end{document}